\documentclass[12pt,epsf,amssymb,qsymbols,axodraw]{article}
\usepackage{tabularx}
\usepackage{array}
\usepackage{graphics}
\usepackage{graphicx}
\usepackage{psfrag}
\usepackage{epsfig}
\usepackage{amsmath}
\usepackage{amssymb}
\usepackage{color}
\usepackage{rotating}
\makeatletter

\usepackage{verbatim}
\newcommand{\bea}{\begin{eqnarray}}
\newcommand{\eea}{\end{eqnarray}}
\newcommand{\beq}{\begin{equation}}
\newcommand{\eeq}{\end{equation}}
\newcommand{\gev}{{\rm GeV}}
\newcommand{\mev}{{\rm MeV}}

\newcommand{\pdir}{p\kern -5.2pt\raise 0.2ex\hbox {/}}
\newcommand{\vdir}{v\kern -5.75pt\raise 0.15ex\hbox {/}}
\newcommand{\kdir}{k\kern -5.75pt\raise 0.15ex\hbox {/}}
\newcommand{\epsdir}{\epsilon\kern -5.0pt\raise 0.15ex\hbox {/}}
\newcommand{\bvdir}{\bar{v}\kern -5.75pt\raise 0.15ex\hbox {/}}
\newcommand{\Ddir}{D\kern -7.75pt\raise 0.20ex\hbox {/}}
\newcommand{\Adir}{A\kern -7.75pt\raise 0.20ex\hbox {/}}
\newcommand{\ldir}{l\kern -5.0pt\raise 0.2ex\hbox{/}}
\newcommand{\varepsdir}{\varepsilon\kern -5.5pt\raise 0.15ex\hbox{/}}

\newcommand{\nn}{\nonumber}
\makeatother

\begin{document}
\thispagestyle{empty} 
\begin{flushright}
\begin{tabular}{l}
LPT, Orsay 09-44 \\
LAL  09-97 \\
\end{tabular}
\end{flushright}
\begin{center}
\vskip 3.0cm\par
{\par\centering \textbf{\LARGE  
\Large \bf Soft Photon Problem in Leptonic $B$-decays}}\\
\vskip 1.25cm\par
{\scalebox{.83}{\par\centering \large  
\sc Damir Be\'cirevi\'c$^{a}$, Benjamin Haas$^{a}$ and Emi Kou$^{b}$}}
{\par\centering \vskip 0.75 cm\par}
{\sl 
$^a$Laboratoire de Physique Th\'eorique (B\^at.~210)~\footnote{Laboratoire de Physique Th\'eorique est une unit\'e mixte de recherche du CNRS, UMR 8627.}\\
Universit\'e Paris Sud, Centre d'Orsay, F-91405 Orsay-Cedex, France.\\
\vspace{2.5mm}
$^b$Laboratoire de l'Acc\'el\'erateur Lin\'eaire, IN2P3-CNRS~\footnote{Laboratoire de l'Acc\'el\'erateur Lin\'eaire est une unit\'e mixte de recherche du CNRS, UMR 8607.}\\
Universit\'e Paris Sud, Centre d'Orsay, F-91898 Orsay-Cedex, France.}\\
\vskip1.cm
 
{\vskip 0.35cm\par}
\end{center}

\vskip 0.55cm
\begin{abstract}
We point out at the peculiarity of $B\to \mu\nu_\mu$ decay, namely  the enhancement of the soft photon events which originate 
from the structure dependent part of the   $B\to \mu\nu_\mu\gamma$ amplitude. This may be a dominant source of systematic uncertainty 
and compromise the projected experimental uncertainty on $\Gamma(B\to \mu\nu_\mu)$.  We show that the effect of these soft photons can be controlled if 
the experimental cut on identification of  soft photons is lowered and especially if the better resolution in identifying the momentum of 
muon emerging from $B\to \mu\nu_\mu$ is made. A lattice QCD computation of the relevant form factors would be highly helpful for 
a better numerical control over the structure dependent soft photon emission.
\end{abstract}
\vskip 2.2 cm 
\setcounter{page}{1}
\setcounter{footnote}{0}
\setcounter{equation}{0}
%%%%%%%%%%%%%%%%%%%%%%%%%%%%%%%%%%%%%%%%
%%%%%%%%%%%%%%%%%%%%%%%%%%%%%%%%%%%%%%%%
%%%%%%%%%%%%%%%%%%%%%%%%%%%%%%%%%%%%%%%%
\noindent

\renewcommand{\thefootnote}{\arabic{footnote}}

\newpage
\setcounter{footnote}{0}
%%%%%%%%%%%  Section 1
\section{Introduction}
One of the most interesting and yet the simplest $B$-meson decays is the leptonic mode $B\to \ell \nu_\ell$, $\ell \in \{ \tau, \mu, e\}$. The expression for its 
decay width is
\bea\label{eq1}
\Gamma(B \to \ell \nu) = { G_\mu^2 \over 8\pi }\times \vert V_{ub}\vert^2 \times m_B^3\left({m_\ell \over m_B}\right)^2\left[  
1-  \left({m_\ell \over m_B}\right)^2 \right] \times  f_B^2\,,
\eea
which we wrote in terms of four factors, of which three are the subject of intense research in the $B$-physics community. The last factor is the decay constant, $f_B$, whose accurate value is still unknown 
although important progress in lattice QCD has been made over the past several years. Whether or not a  percent error on that quantity is achievable in a near future is a topic that is being debated in the lattice community. Present status of the calculation of that quantity on the lattice has been recently reviewed in ref.~\cite{Gamiz}. If the other quantities in eq.~(\ref{eq1}) were known, 
$f_B$ could be extracted from the experimentally measured leptonic decay width which is expected to be done accurately  Super-$B$ factories (Super KEKB and Super-flavour factory)~\cite{superB}.  
Besides, $f_B$ enters
 decisively in the expression for the $B^0_d-\overline B^0_d$ mixing amplitude and its value is essential in understanding the validity of factorization approximation in specific classes of 
 non-leptonic $B$-decays.  The second factor in  eq.~(\ref{eq1}) is  $\vert V_{ub}\vert$ which, together with $\vert V_{td}\vert$, is the smallest entry in the CKM matrix and is very hard to 
 extract by confronting theoretical predictions with experiment. Its value can also be accessed through the inclusive/exclusive semi-leptonic decays. It can also be simply fixed by imposing 
 the CKM unitarity. For that reason 
 this leptonic decay mode is an essential check of consistency when performing the overall fits of the CKM unitarity triangle~\cite{CKM}.
 Finally, in the Standard 
 Model,  the first factor is simply the Fermi constant ($G_\mu=G_F$) which encodes the weak interaction at high energy scales [${\cal O}( m_W)$]. $G_\mu$  can receive appreciable corrections 
 in various extensions of Standard Model. As an example, in ref.~\cite{gino} it has been argued that  this decay mode can be very useful in constraining the charged Higgs mass in the SUSY scenarios 
 of physics beyond standard model with large-$\tan\beta$.  

In short, this channel is very valuable because it can either help fixing $\vert V_{ub}\vert$, or constraining the non-Standard Model physics, or even determining $f_B$ (provided no new physics 
contributes to $G_\mu$, and $\vert V_{ub}\vert$ is determined elsewhere).  This is why this channel is one of the main research targets in the Super-$B$ factories. 

The third factor in  eq.~(\ref{eq1})  exhibits the helicity suppression, so that besides a small CKM coupling,  an extra suppression comes with  $(m_\ell/m_B)^2$, making this process extremely rare 
for $\ell = e,\mu$.  The decay to $\tau$-lepton, although very rare too, is less suppressed and therefore accessible from the $B$-factories. In each of the two $B$-factories the experimenters isolated 
about $20$ events and reported 
\bea
B(B^+\to \tau^+\nu)  =   
\left\{
\begin{array}{lc} 
\left[ 1.79^{+0.56}_{-0.49} ({\rm stat.})^{+0.46}_{-0.51} ({\rm syst.})\right] \times 10^{-4} \,,   &  {\rm Belle}~\cite{BELLE-tau}\ , \\
&  \\
\left[1.8^{+0.9}_{-0.8}\ ({\rm stat.}) \pm 0.4 ({\rm syst.}) \pm 0.2 ({\rm syst.})\right] \times 10^{-4} \,,    &  {\rm BaBar}~\cite{BABAR-tau}\ .
\end{array}
\right.\;
\eea
Since the final $\tau$-lepton is not directly observed but rather reconstructed from its decay products (more specifically, $\tau^-\to e^- \bar \nu_e\nu_\tau$, 
$\mu^- \bar \nu_\mu\nu_\mu$,  $\pi^- \nu_\tau$,  $\pi^0 \pi^- \nu_\tau$),  an irreducible systematic error --due to reconstruction procedure of $\tau$-- diminish the chances to make 
a precision measurement of this decay mode in  Super-$B$ factories. That difficulty is  expected to be circumvented if one was able to observe  $B\to \mu \nu_\mu$ decay directly. 
In this paper we will argue that a new problem emerges in $B\to \mu \nu_\mu$ mode which is peculiar for this decay and is due to the presence of soft photons in the decay product. 
The reason for this phenomenon is related to the fact that the radiative decay lifts the helicity suppression, i.e. it is enhanced by a factor $\propto (m_B/m_\ell)^2$, which is large 
in spite of the suppression by the electromagnetic coupling $\alpha_{\rm em}=1/137.036$~\cite{PDG}.

In what follows we will explain how a large fraction of events that are selected as $B\to \mu \nu_\mu$ might in fact be $B\to \mu \nu_\mu \gamma_{\rm soft}$,  with the soft photon considered as originated from one of the backgrounds of $B\to \mu\nu_{\mu}$.  We will first explain the origin of the problem, point out the hadronic (non-perturbative) origin of the soft photon emission, and then 
discuss how the selected sample of  ``leptonic events" in experiment can be cleaned from those accompanied by a soft photon.

\section{Radiative leptonic decay}
\begin{figure}[t!]
\begin{center}
\epsfig{file=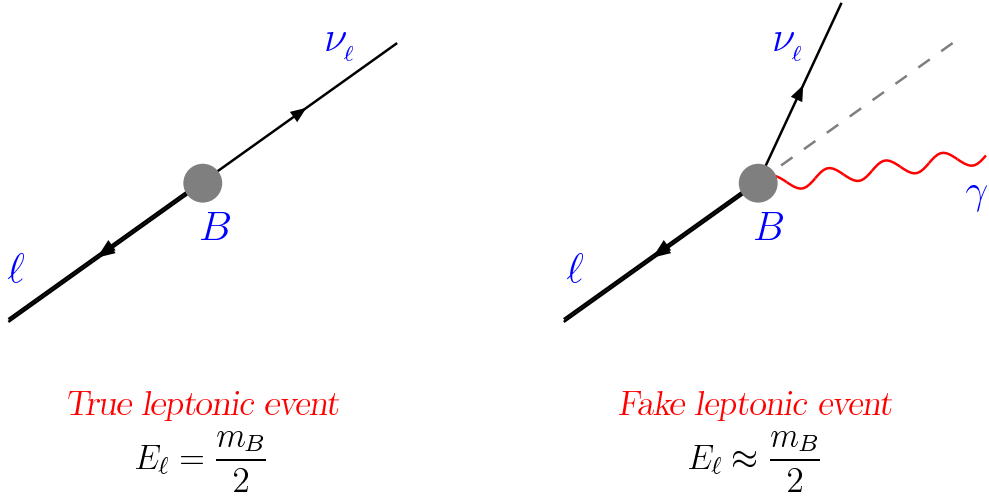, height=6.5cm}
\caption{\label{figA}\footnotesize{\sl 
The pure leptonic decay and the radiative leptonic decay in which the photon is soft and that can be misidentified as a leptonic event. }} 
\end{center}
\end{figure}

\setcounter{equation}{0}
To understand the origin of the problem we remind the reader of the basic elements concerning the radiative $B\to \ell \nu_\ell \gamma$ decay. 
The detailed formulas were derived in several papers, of which we were able to confirm those presented in ref.~\cite{Bijnens}.  The amplitude for this decay   
can be split into three pieces: (i) {\sl inner bremstrahlung} (IB) accounts for the photons emerging from point-like sources (weak vertex, emerging lepton, point-like meson), (ii) 
{\sl structure dependent} (SD) terms, i.e. the photons which probe the internal structure of the decaying meson, and (iii) {\sl interference} (INT) of IB and SD.  It is convenient to work in the $B$ rest frame  and define the variables 
\bea
x=\frac{2E_\gamma}{m_B}\,, \quad  y=\frac{2E_\ell}{m_B}\, ,
\eea
and the angle between $\gamma$ and $\ell$ ($\theta _{\ell\gamma }$) which is related to $x$ and $y$, after setting $p_\nu^2=0$, through~\footnote{
The physically accessible values for $x$ and $y$ from the radiative leptonic decays are: 
\bea
2 r_\ell \leq y \leq 1+ r_\ell^2\,,\quad {\rm and}\quad 1 -\frac{1}{2}  \left( y + \sqrt{y^2 - 4 r_\ell^2}\right) \leq x \leq 1 -\frac{1}{2}  \left( y - \sqrt{y^2 - 4 r_\ell^2}\right) \,.\nn
\eea }
\bea
x=\frac{1}{2} { \left( 2-y+\sqrt{ y^2-4 r_\ell^2 } \right) \left( 2-y-\sqrt{y^{2}- 4 r_\ell^2 }\right) \over 2-y+\sqrt{y^{2}-4r_\ell^2 } \, \cos \theta _{\ell\gamma }}\,,
\eea
where $r_\ell = m_\ell/m_B$. 
The differential decay rate for each of the pieces mentioned above reads:
\bea\label{dGamma}
{\phantom{\huge{l}}}\raisebox{-.6cm}{\phantom{\huge{j}}}
&&{1\over  \Gamma(B\to \ell\nu)}\ { d^2\Gamma^{\rm IB}(B\to \ell\nu\gamma)\over dx\ dy}=
\frac{\alpha_{\rm em} }{ 2 \pi (1-r_\ell^2)^2}  f_{\rm IB}(x,y)  \,, \label{eq:5} \cr
&&{1\over  \Gamma(B\to \ell\nu)}\ { d^2\Gamma^{\rm SD}(B\to \ell\nu\gamma)\over dx\ dy}=\frac{\alpha_{\rm em} }{ 8 \pi r_\ell^2 (1-r_\ell^2)^2}  {m_B^2\over f_B^2}
  \biggl\{  \left[ F_V(x)+F_A(x)\right]^2 f^+_{\rm SD}(x,y)  \biggr.\cr
&& \hspace*{70mm} +\ \biggl. \left[ F_V(x)-F_A(x)\right]^2 f^-_{\rm SD}(x,y)  \biggr\}\,,   \cr
&&{1\over  \Gamma(B\to \ell\nu)}\ { d^2\Gamma^{\rm INT}(B\to \ell\nu\gamma)\over dx\ dy}=\frac{\alpha_{\rm em} }{ 2 \pi (1-r_\ell^2)^2}   {m_B\over f_B}
 \biggl\{  \left[ F_V(x)+F_A(x)\right] f^+_{\rm INT}(x,y)  \biggr.\cr
&& \hspace*{70mm} +\ \biggl. \left[ F_V(x)-F_A(x)\right] f^-_{\rm INT}(x,y)  \biggr\}  \,,
\eea
where we obviously  included the  interference term ``INT". The explicit form of the functions on the right hand side is :
\bea
f_{\rm IB}(x,y)&=&\frac{(1-y+r_\ell^2)}{x^2(x+y-1-r_\ell^2)}\left[ x^2+2(1-x)(1-r_\ell^2)-\frac{2xr_\ell^2(1-r_\ell^2)}{x+y-1-r_\ell^2}\right] \,, \label{eq:12}\\
f_{\rm SD}^+(x,y)&=& (x+y-1-r_\ell^2)\left[(x+y-1)(1-x)-r_\ell^2\right] \,, \\ 
f_{\rm SD}^-(x,y)&=&(1-y+r_\ell^2)\left[(1-x)(1-y)+r_\ell^2 \right]\,, \\
f_{\rm INT}^+(x,y)&=&\frac{1-r+y}{x(x+y-1-r_\ell^2)}\left[(1-x)(1-x-y)+r_\ell^2\right]\,,  \\ 
f_{\rm INT}^-(x,y)&=&\frac{1-y+r_\ell^2}{x(x+y-1-r_\ell^2)}\left[x^2-(1-x)(1-x-y)-r_\ell^2\right]\,.  
\eea
Information about the  meson  structure is encoded in the form factors $F_V(x)$ and $F_A(x)$ which parameterize the matrix element 
\bea
\frac{m_B}{\sqrt{4\pi\alpha_{\rm em} }}\langle \gamma | \bar{b}\gamma_\mu (1-\gamma_5)u| B \rangle= 
F_V(x)\epsilon_{\mu\nu\alpha\beta}\eta^{\nu} p_B^{\alpha}p_{\gamma}^{\beta} +iF_A(x)\biggl[\eta_{\mu}(p_B\cdot p_\gamma )-p^\gamma_\mu (p_B\cdot \eta )\biggr]\,,
\eea
where $\eta$ is the photon polarization vector. The problem that we are emphasizing in this paper is that 
in realistic situations in which one wants to measure accurately the leptonic decay $B\to \mu \nu_\mu$ or $B\to e \nu_e$, many events from the sample are likely to originate from 
$B\to \mu \nu_\mu \gamma$ or $B\to e\nu_e\gamma$ with the photon coming from the SD part of the radiative decay amplitude. 
This may result in a large systematic error on $\Gamma( B\to \mu/e \nu)$ and should be studied carefully. We illustrate this problem in fig.~\ref{figA}.
If one is not able to {\em experimentally} distinguish the events with moderately soft photons then an accurate computation of the $F_{V,A}(x)$ form factors is necessary  
because only in that way the systematic error due to the events accompanied by those photons can be kept under control.
The computation of  $F_{V,A}(x)$ is, however, more complicated a problem than computing  the decay constant $f_B$ itself, and  
this would seriously compromise our chances to extract  $\vert V_{ub}\vert$, or to search/test the presence of  new physics via leptonic $B$-decays.
To illustrate this problem on more quantitative ground we will first estimate the form factors  $F_{V,A}(x)$ in the soft photon region and then discuss their 
impact on leptonic decays from the Dalitz plot considerations.

\section{Form Factors}
Before we discuss the integration over $x$ and $y$ in eq.~(\ref{dGamma}), we should provide an estimate for the form factors $F_{V,A}(x)$. 
In the region  close to $x\to 0$, in which the photon is  soft [$q^2 \to q^2_{\rm max}$], it is reasonable to assume the nearest pole dominance.  
It consists in replacing $\langle \gamma | \bar{b}\gamma_\mu (1-\gamma_5)u| B \rangle \to  \langle \gamma | V_\mu - A_\mu | B \rangle^{\rm pole}$. 
We first discuss the vector current matrix element:
\bea
\frac{m_B}{\sqrt{4\pi\alpha_{\rm em} }}  \langle \gamma | \bar{b}\gamma_\mu  u  | B \rangle^{\rm pole} & =& \frac{m_B}{\sqrt{4\pi\alpha_{\rm em} }}  \sum_\lambda {\langle 0\vert  \bar u \gamma_\mu b\vert B^\ast(\epsilon_\lambda)\rangle \langle 
B^\ast(\epsilon_\lambda)\vert B \gamma\rangle \over q^2 - m_{B^\ast}^2}\,,
\eea
so that, after using the standard definitions, 
\bea
\langle 0\vert  \bar u \gamma_\mu b\vert B^\ast(\epsilon_\lambda)\rangle  &=& \epsilon^\lambda_\mu m_{B^\ast} f_{B^\ast}\,,\cr
 \langle \gamma(p_\gamma,\eta_{\lambda'})B(p_B)\vert B^\ast(q,\epsilon_\lambda)\rangle  &=& e\ \varepsilon_{\mu\nu\alpha\beta} \  
\eta_{\lambda'}^\mu{\epsilon_{\lambda}}^{\nu} q^\alpha p_B^\beta  \ g_{B^\ast B\gamma}\,,
\eea
we obtain 
\bea
F_V(q^2) = {f_{B^\ast } m_B  g_{B^\ast B\gamma}\over m_{B^\ast}} {1 \over 1- q^2/m_{B^\ast}^2}\,.
\eea
Now, by replacing $q^2=m_B^2(1 - x)$, the form factor in terms of $x$-variable  reads
\bea\label{fv}
F_V(x) = {C_V \over x-1+\Delta_b}\,,\quad {\rm with}\,\;\; C_V={m_{B^\ast}\over m_B}f_{B^\ast} g_{B^\ast B\gamma}\,,\; \Delta_b={m_{B^\ast}^2\over m_B^2}\,.
\eea
In this form the physics problem becomes more apparent because it shows that for the soft photon, 
$x=2E_\gamma/m_B\to 0$, the form factor $F_V(x) $ becomes nearly divergent, which is a consequence of the fact that the nearest pole (vector meson $B^\ast$) is very close to 
the pseudoscalar meson. Numerically,  $1- \Delta_b=-0.017$. Such a phenomenon is much less  relevant in charm physics where $\Delta_c=1.157$, and it is practically negligible in  
kaon physics where $\Delta_s=3.262$. 

Similarly, for the axial current we have
\bea
\frac{m_B}{\sqrt{4\pi\alpha_{\rm em} }}  \langle \gamma | \bar{b}\gamma_\mu  \gamma_5 u  | B \rangle^{\rm pole} & =& \frac{m_B}{\sqrt{4\pi\alpha_{\rm em} }}  \sum_\lambda {\langle 0\vert  \bar u \gamma_\mu\gamma_5  b\vert B_1^\prime (\epsilon_\lambda)\rangle \langle 
B_1^\prime (\epsilon_\lambda)\vert B \gamma\rangle \over q^2 - m_{B_1^\prime}^2}\,.
\eea
Since the definition of the relevant coupling to a  soft photon  is not very standard and since various definitions are employed in the literature, we now briefly explain how we defined it. 
Starting from the matrix element of the light electromagnetic current, and by writing $P=p_1+p$, $p_\gamma=p_1-p$
\bea
\langle B(p)\vert J^\mu \vert B_1^\prime (p_1)\rangle = f_1(p_\gamma^2) (P\cdot p_\gamma)\epsilon_\lambda^\mu + (p \cdot \epsilon^\lambda) \left[ f_2(p_\gamma^2) P^\mu + f_3(p_\gamma^2) p_\gamma^\mu \right]\,.
\eea
By imposing the transversity of the on-shell photon ($p_\gamma^2=0$), the two form factors become related at $p_\gamma^2=0$, i.e. 
$f_1(0)(\epsilon\cdot p_\gamma) = - f_2(0) (\epsilon\cdot p)$.  Finally, after saturating by the  photon polarization vector we get 
\bea
 \langle \gamma(p_\gamma,\eta_{\lambda'})B(p)\vert {B_1^\prime }(p_1,\epsilon_\lambda)\rangle &=&
\lim_{p_\gamma^2\to 0}\eta_{\lambda^\prime}^{\ast\mu} \langle B(p)\vert J_\mu \vert B_1^\prime (p_1)\rangle  \cr
& =&  
  f_1(0)  \left[ (\epsilon\cdot \eta^\ast)(P\cdot p_\gamma) - (\epsilon\cdot p_\gamma) (P\cdot \eta^\ast)\right] \cr
  &=& e \underbrace{ (2/e) f_1(0) }_{ {\displaystyle{g_{B_1^\prime B\gamma} } }} \left[ (\epsilon\cdot \eta^\ast)(p\cdot p_\gamma) - (\epsilon\cdot p_\gamma) (p\cdot \eta^\ast)\right]\,.
\eea
Together with $\langle 0\vert  \bar u \gamma_\mu \gamma_5 b\vert  {B_1^\prime} (\epsilon_\lambda)\rangle  = i \epsilon^\lambda_\mu m_{B_1^\prime} f_{B_1^\prime}$, we then have~\footnote{Notice that with the above definitions of $g_{B^\ast B\gamma}$ and  $g_{B_1^\prime B\gamma}$ the electromagnetic decay widths read
\bea
\Gamma(B^\ast \to B\gamma)={\alpha_{\rm em}\over 3}g_{B^\ast B\gamma}^2 \left( { m_{B^\ast}^2 - m_B^2 \over 2 m_{B^\ast} }\right)^3\,,\quad
\Gamma(B_1^\prime \to B\gamma)={\alpha_{\rm em}\over 3}g_{B_1^\prime B\gamma}^2 \left( { m_{B_1^\prime}^2 - m_B^2 \over 2 m_{B_1^\prime} }\right)^3\,.\nn
\eea}
\bea
F_A(q^2) & =& {f_{B_1^\prime} m_B  g_{B_1^\prime B\gamma}\over m_{B_1^\prime}} {1 \over 1- q^2/m_{B_1^\prime}^2} \,.
\eea
In terms of $x$-variable,
\bea\label{fa}
F_A(x) = {C_A \over x- 1+\overline \Delta_b}\,,\quad {\rm with}\,\; C_A={m_{B_1^\prime}\over m_B}f_{B_1^\prime} g_{B_1^\prime B\gamma}\,,\; \overline \Delta_b={m_{B_1^\prime}^2\over m_B^2}\,.
\eea
From the recent experimental observation, $m_{B_1^\prime} - m_B =441.5 \pm 2.7$~\cite{PDG}, we get $\overline \Delta_b=1.174(1)$. In other words, $| 1- \overline \Delta_b| \gg |1- \Delta_b|$, and it regularizes 
the axial form factor in the small $x$ region. The impact of the vector form factor is therefore far more important than that of the axial form factor. 

We should now fix the ``residua" $C_{V,A}$. To that end we need an estimate for the couplings  $g_{B^\ast B\gamma}$ and $g_{B_1^\prime B\gamma}$, and for the decay 
constants  $f_{B^\ast}$ and $f_{B_1^\prime}$. Concerning the decay constant $f_{B^\ast}$, from the (averaged) lattice QCD estimate $f_{B^\ast}/f_B=1.03(2)$~\cite{fB*}, together with $f_B=195(11)$~MeV~\cite{Gamiz} we have $f_{B^\ast}=201(12)$~MeV. As for the axial decay constant, to our knowledge, no lattice QCD determination of its value has been made so far. The model of ref.~\cite{dunietz} gives $f_{B_1^\prime}=206(29)$~MeV, which covers the values obtained  by using the QCD sum rules in the static heavy quark limit (without radiative corrections), $f_{B_1^\prime}/f_B\approx 1.2$~\cite{sumrules-B1}, and those obtained by using the covariant model, $f_{B_1^\prime}/f_B\approx 0.9$~\cite{fB1-model2}.

Regarding the couplings  $g_{B^\ast B\gamma}$ and $g_{B_1^\prime B\gamma}$, we will estimate their values by using various quark models. 
In the quark model picture, $g_{B^\ast B\gamma}$ is the sum of magnetic moments of the valence quarks, so that in the static heavy quark limit we
 simply  have  $g_{B_q^\ast B_q\gamma} = e_q \mu_V$, where we now specify the light quark, i.e. its charge $e_q$. Since in our case we are interested 
 in the charged $B$-meson,  the light quark is $u$ and  $g_{B^\ast B^\pm \gamma} = (2/3) \mu_V$. As a side remark we see that when discussing the charged 
 charmed mesons in the same limit,  $g_{D^\ast D^\pm \gamma} = -(1/3) \mu_V$,  and therefore the effect of soft photons will  be suppressed not only because 
 $ \Delta_c >  \Delta_b$, as discussed above, but also because the soft photon coupling is halved with respect to the $B$-meson case. 
Similarly, $g_{B_1^\prime B^\pm \gamma}=(2/3) \mu_A$, so that the whole question is reduced to finding the values of $\mu_{V,A}$. 
To get an estimate of these couplings we can use the quark model of ref.~\cite{alain}. In the same notation as the one given in that paper we obtain 
\bea
\mu_V&=&{\sqrt{2}\over |p_\gamma |}\int d\vec r\ \overline \Psi_B (r) \vec \gamma\cdot \vec \eta \ \Psi_{B^\ast}(r) \ e^{-i \vec p_\gamma \vec r}\cr
&=&   {2\over 3} \int_0^\infty dr \ r^3 \left[ f_{1/2}^{(-1)\ast }(r)  g_{1/2}^{(-1) }(r) + g_{1/2}^{(-1)\ast}(r)  f_{1/2}^{(-1) }(r) \right]\,,
\eea
and similarly 
\bea
\mu_A&=&  {2\over 3} \int_0^\infty dr \ r^3 \left[ f_{1/2}^{(+1)\ast }(r)  g_{1/2}^{(-1) }(r) + g_{1/2}^{(+1)\ast}(r)  f_{1/2}^{(-1) }(r) \right]\,,
\eea
which leads to $\mu_V=1.51\ \gev^{-1}$ ($g_{B^\ast B^\pm \gamma} = 1\ \gev^{-1}$) and $\mu_A=1.04\ \gev^{-1}$ ($g_{B_1^\prime B\gamma} = 0.7\ \gev^{-1}$)  
if the same set of parameters is used as in ref.~\cite{alain}. Together with masses and decay constants discussed above, this would lead us to 
\bea
C_V= 0.20(1)\,,\quad C_A=0.16(2)\,.
\eea
Moreover for the coupling of the radially excited state to $B$-meson and the soft photon, we obtain $\mu_V^\prime=0.35\ \gev^{-1}$, and  $\mu_A^\prime=0.40\ \gev^{-1}$, 
which translates in the positive correction to the form factor $F_V(x)$ which ranges from $1\%\div 12\%$, when $x\in(0,0.2]$, i.e. in the region that we are focusing in this paper. 
On the other hand the correction to the axial form factor is more problematic because it is large: for the choice of the model parameters made in ref.~\cite{alain}, it amounts to a correction as high as $+60\%$ ,with respect to the nearest pole dominance. ~\footnote{One of the ingredients in this discussion is the ratio of decay constants. By combining the results of ref.~\cite{fB1-model2} with the masses of radial excitations reported in ref.~\cite{dipierro} $m_{B^\ast}(2S_1)=5.92$~GeV,  $m_{B^\prime_1}(2P_1)=6.19$~GeV, we have $f^\prime_{B^\ast}/f_{B^\ast}\simeq 1.03$, and $f^\prime_{B_1^\prime}/f_{B^\prime_1}\simeq 0.99$.}

We also checked that the change of model parameters only moderately affects the value of  $\mu_V$, whereas the values of $\mu_A^{(\prime)}$ are very sensitive to the  choice of model parameters. 
Unfortunately, to our knowledge, there are no other computations of the axial coupling $g_{B^\prime_1 B \gamma}$ in the literature. Instead there is quite a number of predictions for the coupling 
$g_{B^\ast B^\pm \gamma}$. A list of results obtained  in various (hopefully representative) models is provided in table~\ref{table:compare}. 
The problem in computing the strong couplings from three-point correlation functions in QCD sum rules is  well known~\cite{us3}. An alternative strategy is to access these couplings via 
form factors. From the QCD sum rule computation of the form factors $F_{V,A}(q^2)$ in an external electromagnetic field made in ref.~\cite{sumrule} and extrapolated to $q^2=(m_B/2)^2$, we can extract the values for $C_{V,A}$. We get 
\bea
C_V\approx 0.27\,,\quad C_A\approx 0.24\,.
\eea
A similar result was obtained in the dispersive model of ref.~\cite{melikhov}. 
In what follows we will use 
\bea\label{CVA}
C_V=0.24\pm 0.04\,,\quad C_A= 0.20\pm 0.05\,,
\eea
which we believe is a good compromise. We reiterate that a better determination of these residua or --even better-- of the form factors (preferably by means of the QCD simulations on the lattice) would be very welcome. 
\begin{table}[t!]
\begin{center}
\hspace*{-5.5mm}
{\scalebox{.92}{ \begin{tabular}{|c|c|c|c|}
\hline
{\phantom{\Huge{l}}}\raisebox{-.1cm}{\phantom{\Huge{j}}}
Method & Ref. & $\Gamma(B^{\ast +}\to B^+\gamma)$  [${\rm keV}^{-1}$]  &$g_{B^{\ast +} B^+\gamma}$ [$\gev^{-1}$]   \\
 \hline
{\phantom{\Huge{l}}}\raisebox{-.1cm}{\phantom{\Huge{j}}}
\hspace*{-1cm}{\small\sl Light Front Model} & \cite{choi}  & $0.40(3)$  & $1.32(5)$ \\ 
{\phantom{\Huge{l}}}\raisebox{-.1cm}{\phantom{\Huge{j}}}
  & \cite{jaus}  & $0.43$  & $1.37$\\ 
{\phantom{\Huge{l}}}\raisebox{-.1cm}{\phantom{\Huge{j}}}
\hspace*{-1cm}{\small \sl Chiral Quark Model}&\cite{goity}  & $0.5\div 0.8$   &  $1.5\div 1.9$\\ 
{\phantom{\Huge{l}}}\raisebox{-.1cm}{\phantom{\Huge{j}}}
\hspace*{-.7cm}{\small\sl Schr\"odinger-like Model}(``S")& \cite{galkin}  & $0.24$  &  $1.03$\\ 
\hspace*{3.75cm}(``V")  &  & $0.29$  &  $1.13$\\ 
{\phantom{\Huge{l}}}\raisebox{-.1cm}{\phantom{\Huge{j}}}
\hspace*{-1cm}{\small\sl Salpeter-like Model} & \cite{colangelo}  & $0.24$  & $1.04$\\ 
{\phantom{\Huge{l}}}\raisebox{-.1cm}{\phantom{\Huge{j}}}
\hspace*{-1cm}{\small\sl Bag Model} & \cite{orsland}   & $0.27$  & $1.09$\\ 
{\phantom{\Huge{l}}}\raisebox{-.1cm}{\phantom{\Huge{j}}}
\hspace*{-1cm}{\small\sl Dirac Model} & this work   & $0.23(5)$  & $1.0(1)$\\ 
\hline 
{\phantom{\Huge{l}}}\raisebox{-.1cm}{\phantom{\Huge{j}}}
\hspace*{-.65cm}{\small\sl QCD Sum Rule  } & \cite{aliev}  & $0.63$  & $1.66$\\ 
{\phantom{\Huge{l}}}\raisebox{-.1cm}{\phantom{\Huge{j}}}
\hspace*{-.65cm}  & \cite{zzhu}  & $0.38(6)$  &  $1.3(1)$\\ 
{\phantom{\Huge{l}}}\raisebox{-.1cm}{\phantom{\Huge{j}}}
\hspace*{-.65cm}  & \cite{narison}  & $0.10(3)$  &  $0.65(10)$\\ 
\hline 
\end{tabular} }}
\end{center}
\caption{\label{table:compare}
\footnotesize  The table of results for the radiative $B^\ast$ meson decays as computed by various quark models and QCD sum rules. 
``S" and ``V" label the scalar or vector potential in the model of ref.~\cite{galkin}. }
\end{table}
\section{Dalitz Plot - How to deal with soft photons?}
\begin{figure}
\begin{center}
\psfrag{x}{\hspace*{-14mm}\color{red}\large $x=2 E_\gamma/m_B$}
\psfrag{y}{}
\psfrag{SDm}{\hspace*{-10mm}\color{red}\large $y=2 E_\mu/m_B$}
\psfrag{mu}{\vspace{-15mm}\hspace*{-18mm}\Large $\displaystyle{\quad { \color{red} { 1\over \Gamma(B\to \mu \nu)  }{
 d^2\Gamma(B\to \mu \nu \gamma)\over dx\ dy}} \quad \over {}}$ }
\psfrag{c}{\color{blue}\large $x$}
\epsfig{file=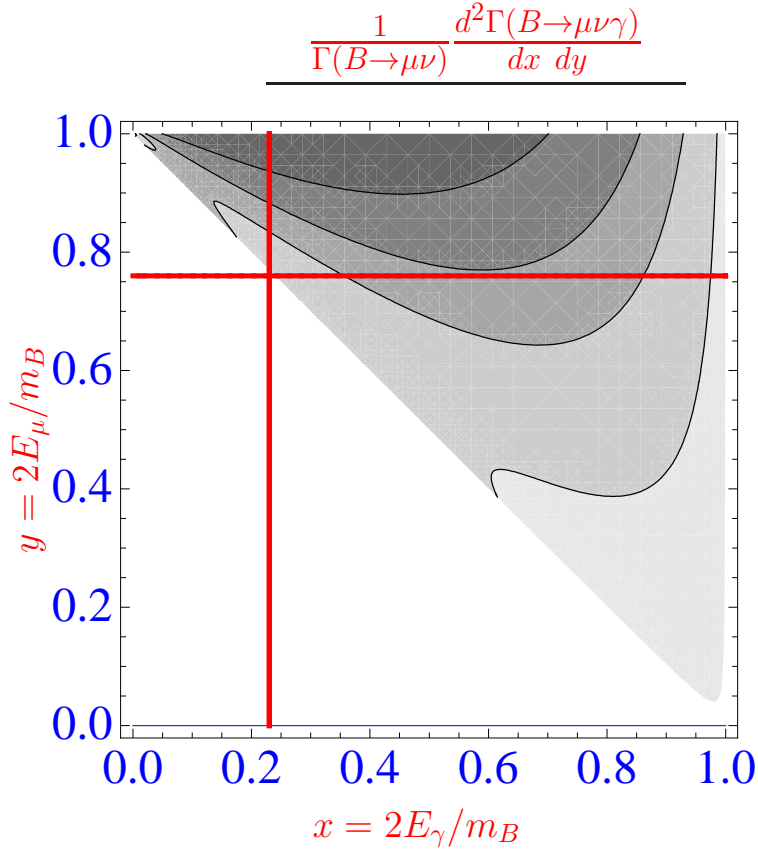, height=10.5cm}
\caption{\label{figB}\footnotesize{\sl 
The structure dependent part of the Dalitz plot of $B\to \mu \nu_\mu \gamma$ decay.  The shaded areas --from bright to dark-- correspond to $1$, $5$, $10$, $20$ and $40$ times  
 the pure leptonic $B\to \mu \nu_\mu$ events.  We also draw the typical cuts which can be realistically employed in experiments while identifying the leptonic decays: the radiative 
 events that are right from the vertical line ($x_{\rm cut}$) are properly taken care of, as well as those below the horizontal line  ($y_{\rm cut}$).}} 
\end{center}
\end{figure}

We are now in a position to estimate the amount of events that can be misidentified in experiment as if they were leptonic while they are actually the 
radiative leptonic decays. In fig.~\ref{figB} we show the part of the Dalitz plot which is due to  non-perturbative structure dependent (SD) part of 
$B\to \mu \nu_\mu \gamma$  decay amplitude. To produce that plot we used the form factor~(\ref{fv},\ref{fa}) with the numbers quoted in eq.~(\ref{CVA}). 
To explain the lines which denote the experimental cuts we should briefly remind the reader about the selection criteria for the true $B\to \mu\nu_\mu$ event. 
The $B\to \mu\nu_\mu$ has a very tight kinematics, namely a muon ($\mu$) and a missing energy ($\nu_\mu$), both carry exactly a half of the $B$-meson momentum ($m_B/2$). In reality, however, the initial $B$-meson is not produced exactly at rest while its momentum can not be measured very precisely. 
As a result, the actual event selection is done by allowing a slightly loosened kinematics:  $\mu$ and $\nu_\mu$ carrying  a momentum  within the range of $m_B/2\ \pm$ a few hundred MeV. 
This momentum ambiguity of the initial state, of course, occurs on the tagging side of the $\overline{B}$ meson which is produced together with $B$ at the $e^+e^-$ collision. The trouble begins 
when an additional neutral particle (without a clear track) is observed with a momentum less than this initial state ambiguity:  the mass reconstruction of $B$ and $\overline{B}$ may both be able to accommodate such  particle. 
That means, one can not distinguish the true event of $B\to \mu\nu_\mu$ with a $\gamma_{\rm soft}$ from the $\overline{B}$ decay and the false event of $B\to \mu\nu_\mu\gamma_{\rm soft}$. 
In order not to miss the former type of the event, the leptonic decay selection criteria has to be further modified, namely by selecting the events with $\mu$ with energy 
$m_B/2\ \pm$ a few hundred MeV {\it plus} allowing photons with energy less than a few hundred MeV.  But then, the $B\to \mu\nu_\mu\gamma_{\rm soft}$ events that are situated in the left-upper 
corner of the Dalitz plot in fig.~\ref{figB}   perfectly pass the $B\to \mu\nu_\mu$ selection criteria.    

In general, allowing to have extra soft photons in the signal is appropriate since the $\gamma_{\rm soft}$ from the $\overline{B}$  occurs very frequently while 
those coming from  $B\to \mu\nu_\mu\gamma_{\rm soft}$ are  $\alpha_{\rm em}$-suppressed. 
However, in this particular case, this argument breaks down since that suppression is largely compensated by the chiral enhancement factor of $B\to \mu\nu_\mu\gamma$,  $m_B/m_\mu$, with respect to $B\to \mu\nu_\mu$. 
In addition, since most of the SD radiative decays occur in the upper end of the Dalitz plot as we show in fig.~\ref{figB}, the false events could be sizable. The ideal solution to this problem 
would involve a full-reconstruction of the $\overline{B}$ side, so that all extra-photons in the event would be forbidden.~\footnote{Indeed such an analysis is performed by using the  exclusive hadronic mass reconstruction on the $\overline{B}$ side. The recent study by the Babar collaboration seems to be encouraging in the sense that  the photon cut can be substantially lowered.  However, the full mass reconstruction of $\overline{B}$ is lacking in most of the analyses since it entails a considerable loss in statistic. } Alternative solution, which we discuss here, is to estimate the false event and subtract them away. Notice that a discussion on the issue of soft photons is lacking  in all the preliminary studies of the feasibility of precision measurement of $B\to \mu \nu_\mu$ decay rate.

In order to estimate the number of $B\to \mu\nu_\mu\gamma$ event which pass the event selection criteria for the leptonic decay, we need precise values of the energy cut for muon and for the extra photon. 
These cuts are imposed differently in each experiment~\cite{BELLE-tau,BABAR-tau,BABAR-mu,BELLE-gamma}. To illustrate the amount of associated systematic uncertainty, here  we chose various 
values of these cuts. 
The identification of the prompt muon, as mentioned above, is bound to an ambiguity  of a few hundreds MeV, which is indicated by the horizontal line in fig.~\ref{figB}.
The vertical line in fig.~\ref{figB} represents the photon energy cut: the cut on photons in this situation means a distinction between the photons that are identified to be coming from $B\to \mu\nu_\mu\gamma$  and are subtracted away (experimentally), and those that are  below $E_\gamma^{\rm cut}$ and selected as if they were leptonic events. 

\begin{figure}[h!]
\begin{center}
\psfrag{Ena22gev}{\hspace*{1cm}\color{blue} ${\mathtt{E_\mu^{\rm cut}=2.2\  \gev}}$}
\psfrag{Ena24gev}{\hspace*{1cm}\color{blue} ${\mathtt{E_\mu^{\rm cut}=2.4\  \gev}}$}
\psfrag{Ena26gev}{\hspace*{1cm}\color{blue} ${\mathtt{E_\mu^{\rm cut}=2.6\  \gev}}$}
\psfrag{Ecut}{\hspace*{2.4cm}$E_\gamma^{\rm cut}$}
\psfrag{SDm}{$R(E_\gamma^{\rm cut}, E_\mu^{\rm cut})$}
\psfrag{mu}{\vspace{-15mm}\hspace*{-18mm}\Large $\displaystyle{\quad { \color{red} { 1\over \Gamma(B\to \mu \nu)  }{
 d^2\Gamma(B\to \mu \nu \gamma)\over dx\ dy}} \quad \over {}}$ }
\psfrag{c}{\color{blue}\large $x$}
\epsfig{file=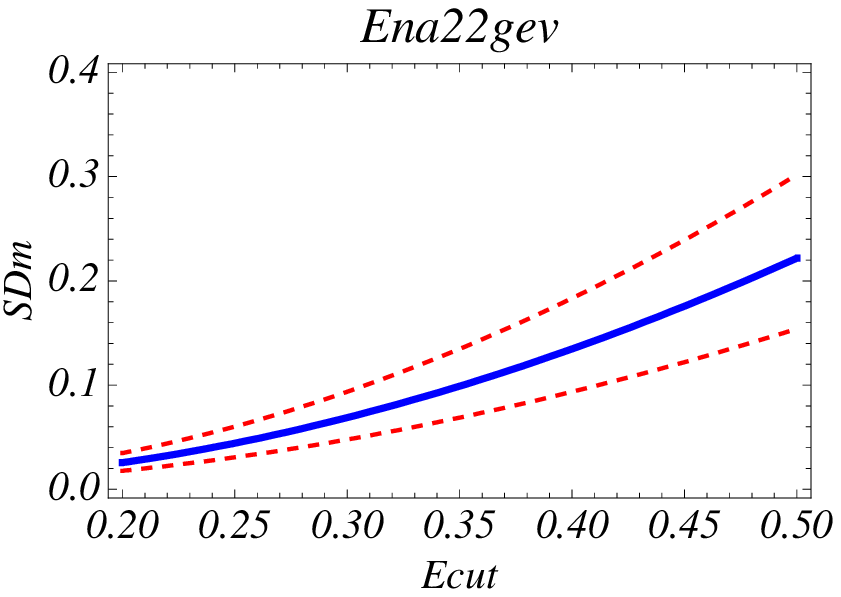, height=5.8cm}\\ 
{\phantom{\Huge{l}}}\raisebox{-.1cm}{\phantom{\Huge{j}}}
\epsfig{file=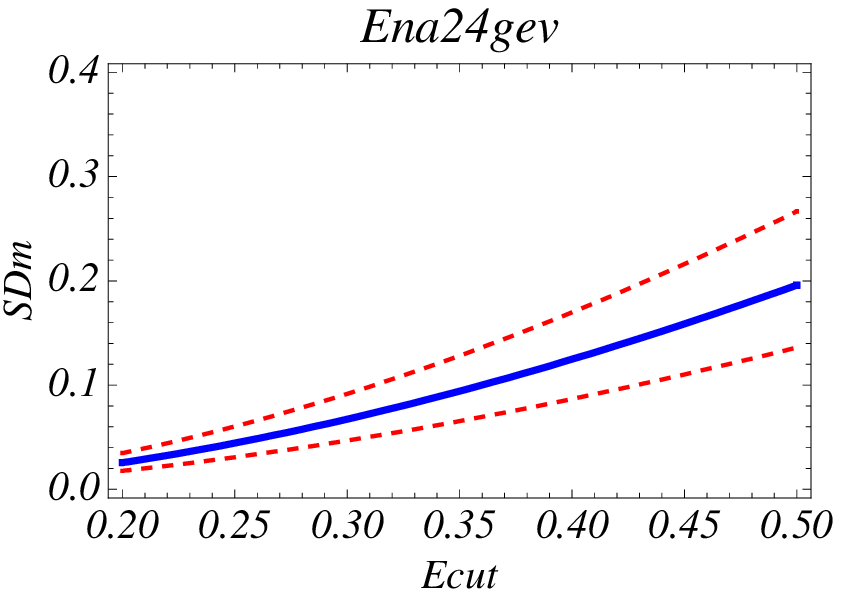, height=5.8cm}\\
{\phantom{\Huge{l}}}\raisebox{-.1cm}{\phantom{\Huge{j}}}
\epsfig{file=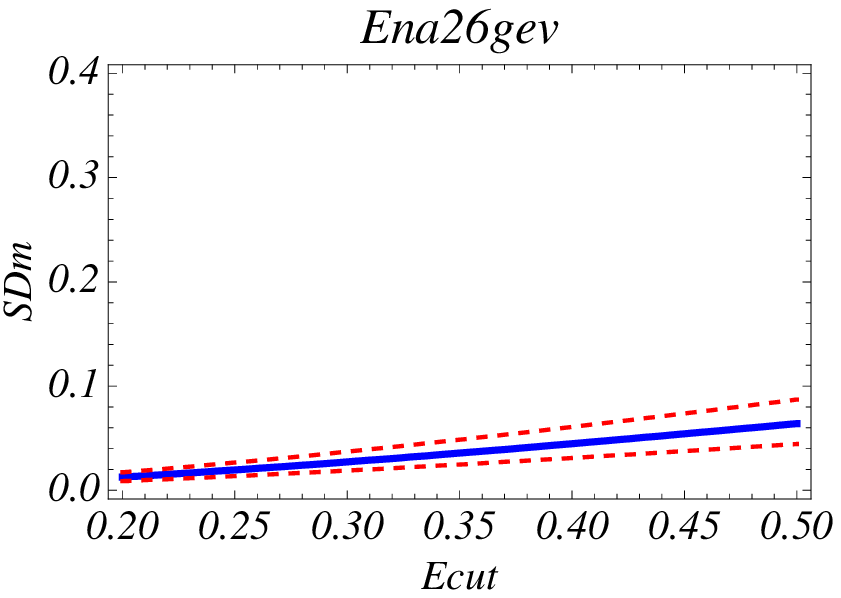, height=5.8cm}
\caption{\label{figC}\footnotesize{\sl 
The amount of the soft photon radiative leptonic events with respect to the leptonic decay as a function of the soft photon cut, and for three different  fixed values of $E_\mu^{\rm cut}$.  The thick curve correspond to the central values for the form factors $F_{V,A}(x)$ while the dashed lines are correspond to the error bars. $R(E_\gamma^{\rm cut}, E_\mu^{\rm cut})= 
\Gamma^{SD}(B\to \mu \nu\gamma; E_\gamma < E_\gamma^{\rm cut}; E_\mu > E_\mu^{\rm cut})/ \Gamma(B\to \mu \nu)$.}} 
\end{center}
\end{figure}

 In fig.~\ref{figC} we show the error made on the leptonic decay width due to the SD soft photons only as a function of the photon energy cut 
$x_{\rm cut}\in (0.075, 0.2)$, i.e. $E_\gamma^{\rm cut}\in (200, 500)$~MeV.
To that purpose we considered the ratio 
\bea\label{RE}
R(E_\gamma^{\rm cut}, E_\mu^{\rm cut})&=&{\Gamma^{SD}(B\to \mu \nu\gamma; E_\gamma < E_\gamma^{\rm cut}; E_\mu > E_\mu^{\rm cut})\over  \Gamma (B\to \mu \nu)} \cr
&=& \int_0^{x_{\rm cut}}dx\int_{y_{\rm cut}}^1 dy \ {1\over  \Gamma(B\to \ell\nu)}\ { d^2\Gamma^{\rm SD}(B\to \ell\nu\gamma)\over dx\ dy} \,,
\eea
where the function under the integral in given in the second line of eq.~(\ref{dGamma}). The range  $E_\gamma^{\rm cut}\in (200,500)$~MeV  includes a realistic values in  
the future experimental data analyses. The price to pay when lowering $E_\gamma^{\rm cut}$, however, is a considerable loss in statistics which then worsens a targeted 
experimental accuracy in $\Gamma(B\to \mu\nu_\mu)$. 
What we also observe is that $R(E_\gamma^{\rm cut}, E_\mu^{\rm cut})$ depends quite substantially on the cut in lepton energy. In fig.~\ref{figC} the illustration is provided for  
three choices $E_\mu^{\rm cut} = 2.2$~GeV, $ 2.4$~GeV, and  $2.6$~GeV, from which we read that getting this cut as close to $m_B/2$ as possible may radically reduce the effects of soft photons.
In particular if one can push $E_\mu^{\rm cut} > 2.5$~GeV, the spillover of the radiative leptonic events becomes indeed small. 
Finally, note that in our window of photon energies the dependence on the photon energy cut is nearly linear. 
For the form factors, the values of which we fixed in the previous section, we see that the error on the leptonic decay width due to misidentified leptonic events is about $20\div 30\%$ for reasonable 
choices of the cuts. We stress again that these numbers  are highly dependent on the input form factors $F_{V,A}(x)$. As one can see from table~\ref{table:compare}, there are models which 
predict the coupling $g_{B^\ast B^+\gamma}$ to be much larger than the values we use here. Those models are plausible, and if --for example-- we used the value predicted 
by the model of ref.~\cite{goity}, the effect of soft photons discussed here would be much larger: even for their smallest $g_{B^\ast B^+\gamma}=0.5$ and by chosing 
$E_\mu^{\rm cut}=2.4$~GeV,  the error on leptonic decay due to soft photons would be between $10\div 80\%$, for the photon energy cuts varied between $200\ \mev \leq E_\gamma^{\rm cut}\leq 500$~MeV.  
Therefore this error is potentially a dominant source of systematic uncertainty and it should be studied carefully. On the experimental side the possibilities to reduce the 
photon identification cut as well as to higher the lepton energy cut are more than desirable. On the theory side, instead, a dedicated study of the $F_{V,A}(x)$ form factors by means of lattice 
QCD is needed if one is to have a handle on remaining soft photons. 

Although we are focusing on the SD term in eq.~(\ref{dGamma}), we checked that by integrating the IB and INT pieces in the same range as indicated in eq.~(\ref{RE}), that these two terms are 
indeed much smaller than the one we are discussing here. The reason is that most of the IB events are concentrated on the diagonal of the Dalitz plot [i.e. along $y_{\rm min}$ while varying $x\in (0,1)$].  
Of course the IB and SD  terms become comparable  if we were able to work with $E_\mu^{\rm cut} > 2.6$~GeV, or  $E_\gamma^{\rm cut} \ll 100$~MeV or so.~\footnote{When integrating the Dalitz plot for IB and INT terms we used the infra-red regulator $\varepsilon_\gamma = 30$, or $50$~MeV.}

\subsection{What about the other heavy meson leptonic decays?}
The other heavy meson leptonic decays are essentially not  nearly as much influenced by this problem as the $B\to \mu/e \nu$ mode is. The problem of radiative decays was discussed long ago 
in ref.~\cite{burdman}, and in a less explicit way in ref.~\cite{radiativePAPERS}.  With the form factors $F_{V,A}(x)$ chosen in a way we discussed in this paper, and by integrating over 
the entire Dalitz plot, we obtain that the SD part of $B\to \tau \nu_\tau \gamma$ is less than $1\%$ with respect to the $B\to \tau \nu_\tau$ decay.  
We also checked that the number of the soft photon events $D\to \tau \nu_\tau\gamma$ is  completely negligible with respect to the corresponding pure leptonic decay.  
 Finally, concerning the $D\to \mu\nu_\mu$ mode, we observe a very weak dependence on the soft photon cut and for 
 $E_\mu^{\rm cut}\approx m_D/2 - 200$~MeV, the error due to misidentification of leptonic events, which are actually the radiative leptonic ones, is  $R(E_\gamma^{\rm cut}, E_\mu^{\rm cut}=0.68\ \gev)< 4\%$. 
 That error falls under $2\%$ if the muon identification momentum  is restrained to $E_\mu^{\rm cut}\approx 0.88$~GeV. Since this effect is nevertheless at the percent level, the chances for 
 checking on the lepton-flavor 
 universality from leptonic $B$ and/or $D$ decays, as proposed in ref.~\cite{isidori2}, are thin.

\section{Summary and conclusions}
We summarize our findings as follows:
\begin{itemize}
\item[$\circ$] As any other leptonic decay of a pseudoscalar meson, the $B$-decays also need  a regularization of the soft photon emission from the point-like particles in the inner bremstrahlung amplitude of the radiative $B\to \ell \nu_\ell$ background process. That is usually made by imposing a cut $\varepsilon_\gamma^{\rm cut}$ on the soft photons so that the $x\to 0$ divergence is avoided.  Experimenters 
treat those photons by Monte Carlo (see ref.~\cite{photos} and references therein). Here we focus on the other part of the  $B\to \mu \nu_\mu \gamma$ amplitude, namely the hadronic structure dependent one.
\item[$\circ$] We show that  the  $B\to \mu \nu_\mu$ [and/or $B\to e \nu_e$] decay is peculiar because of the fact that  the emission of soft photons that are not discernible by detectors and which originate 
from the SD   part of the  $B\to \mu \nu_\mu \gamma$ amplitude  amount to uncomfortably large fraction of misidentified $B\to \mu \nu_\mu$ events. 
There are two important reasons for that enhancement: (a) Contrary to $B\to \mu \nu_\mu$, the radiative $B\to \mu \nu_\mu \gamma$ decay is not helicity suppressed and therefore it picks up a factor $\propto (m_B/m_\mu)^2$, which is large in spite of the suppression by $\alpha_{\rm em}$; (b) the structure dependent term involves the hadronic form factors, of which particularly important is the vector form factor because its nearest pole at $B^\ast$ is very close to $m_B$, making the form factor increasing abruptly in the soft photon region, i.e. of small $x$. 
\item[$\circ$] By a simple model estimates of the residua of the form factors, we show that in realistic situations the systematic error on identification of the leptonic  
$B\to \mu \nu_\mu$ decay, only due to these SD soft photons, is about $20\%$ for the present experimental set-up. We also show that lowering the cuts on photon energy $E^{\rm cut}_\gamma$, 
and especially a refinement of the momentum identification of the emerging muon, can considerably reduce this effect. 
\item[$\circ$] This problem has not been treated so far and the projected uncertainty on $B(B\to \mu\nu_\mu)$ did not take into account the effect of SD soft photons~\cite{superB}, 
which --as we just argued-- can be overwhelmingly large. The current ideas on how to lower $E^{\rm cut}_\gamma$ may partly  be helpful although their implementation results in a considerable loss 
of statistics~\cite{schwert}.
\item[$\circ$] Our rough $20\%$ estimate should be refined. A model independent computation of the form factors $F_{V,A}(x)$ on the lattice would be extremely helpful in keeping these soft photon 
effects under control.
\end{itemize}

\section*{Acknowledgements}
We thank  the partial support of `Flavianet'  (EU contract MTRN-CT-2006-035482), and of  the ANR (contract “DIAM” ANR-07-JCJC-0031). The work of E.K. was supported by the European Commission Marie Curie Incoming International Fellowships under the contract MIF1-CT-2006-027144 and by the ANR (contract "LFV-CPV-LHC" ANR-NT09-508531). Discussions with S.Robertson, C.~Schwanda and A.~Stocchi are kindly acknowledged too.

\end{document}